\documentclass[superscriptaddress,floatfix,longbibliography]{revtex4-2}
\usepackage{amsmath,amssymb} % math symbols
\usepackage{bm} % bold math font
\usepackage{graphicx} % for figures
\usepackage{comment} % allows block comments
\usepackage{enumitem}
\setlist{noitemsep,leftmargin=*,topsep=0pt,parsep=0pt}
\usepackage{xspace}
\usepackage{xcolor} % \textcolor{red}{text} will be red for notes
\definecolor{lightgray}{gray}{0.6}
\definecolor{medgray}{gray}{0.4}
\usepackage{hyperref}

\usepackage{comment}
\usepackage{siunitx}
\usepackage{tabularx}
\usepackage{graphicx}

\usepackage[T1]{fontenc}
\newcommand\ket[1]{\vert{#1}\rangle}
\newcommand\bra[1]{\langle{#1}\vert}
\hypersetup{
colorlinks=true,
urlcolor= blue,
citecolor=blue,
linkcolor= blue,
}

\begin{document}

\title{Entanglement of edge modes in (very) strongly correlated topological insulators}
%Detection of topological phases using entanglement entropy}
\author{Nisa Ara}
\email{p20210035@goa.bits-pilani.ac.in}
\email{p20210036@goa.bits-pilani.ac.in}
\author{Rudranil Basu}
\author{Emil Mathew}
\email{rudranilb@goa.bits-pilani.ac.in}
\author{Indrakshi Raychowdhury}
\email{indrakshir@goa.bits-pilani.ac.in}

\affiliation{Department of Physics, Birla Institute of Technology and Science Pilani, Zuarinagar, Goa 403726, India}\affiliation{Center for Research in Quantum Information and Technology, Birla Institute of Technology and Science Pilani, Zuarinagar, Goa 403726, India}
%\affiliation{, Birla Institute of Technology and Science Pilani, Zuarinagar, Goa 403726}
\bibliographystyle{unsrt}

\bibliographystyle{unsrt}

\begin{abstract}
Identifying topological phases for a strongly correlated theory remains a non-trivial task, as defining order parameters, such as Berry phases, is not straightforward. Quantum information theory is capable of identifying topological phases for a theory that exhibits quantum phase transition with a suitable definition of order parameters that are related to different entanglement measures for the system. In this work, we study entanglement entropy for a bi-layer SSH model, both in the presence and absence of Hubbard interaction and at varying interaction strengths. For the free theory, edge entanglement acts as an order parameter, which is supported by analytic calculations and numerical (DMRG) studies. We calculate the symmetry-resolved entanglement and demonstrate the equipartition of entanglement for this model which itself acts as an order parameter when calculated for the edge modes. As the DMRG calculation allows one to go beyond the free theory, we study the entanglement structure of the edge modes in the presence of on-site Hubbard interaction for the same model. A sudden reduction of edge entanglement is obtained as interaction is switched on. %, and interaction strength lies even in the perturbative regime. 
The explanation for this lies in the change in the size of the degenerate subspaces in the presence and absence of interaction. We also study the signature of entanglement when the interaction strength becomes extremely strong and demonstrate that the edge entanglement remains protected. In this limit, the energy eigenstates essentially become a tensor product state, implying zero entanglement. However, a remnant entropy survives in the non-trivial topological phase which is exactly due to the entanglement of the edge modes.

%We study von Neumann entropy as a non-local order parameter for the coupled SSH model in opposite dimerization (AB stacking). We investigate symmetry-resolved entanglement across topological phase transition in the SSH model as well as the coupled system. With DMRG, we show the effects of on-site interaction in the system.
\end{abstract}

\maketitle
\tableofcontents

\section{Introduction}

For a certain class of non-interacting fermionic theories, the existence of non-trivial topological phases is associated with symmetry structures as well as the dimensionality of the model, which is properly classified in the literature\cite{kitaev2009periodic,schnyder2008classification,altland1997nonstandard}.
The signature of a non-trivial phase is captured via the existence of edge excitations \cite{fu2007topological,fu2007topologicala,moore2007topological,qi2011topological,hasan2010colloquium} and quantified by calculating topological order parameters like a Zak phase for the (one-dimensional) system \cite{zak1989berry}.
%\inote{@NISA: add definition of Zak mode along with expression here.} 
The Zak phase is the berry phase acquired by the occupied energy moving bands across the Brillion zone and can be expressed as  $\Phi= \int_{\mathrm{BZ}}\langle X_k| \partial_k |X_k \rangle\,dk$ where $|X_k \rangle$ is the quasimomentum $k$ dependent eigenvector of the Hamiltonian (kernel). 

Albeit, topological insulators have been classified \cite{tang2012interacting,fidkowski2011topological} even in the presence of interaction, defining an order parameter is still not well understood. In the Su-Schrieffer-Heeger (SSH) model \cite{heeger1988solitons,su1979solitons,atala2013direct}, the prototypical example of 1-dimensional topological insulators, two distinct topological phases are attributed due to staggered hopping amplitudes. In the non-trivial topological phase, there are two zero energy edge modes. 
\textcolor{black}{Motivated by the fact that the existence of edge modes is a prominent signature of a topologically non-trivial phase, the entanglement entropy of edge modes has been studied in SSH or similar models extensively in the presence of interactions as well \cite{wang2015detecting,micallo2020topological,verbin2013observation,kraus2012topological,ryu2006entanglement,ye2016entanglement,tan2020detection,zhou2023exploring,sirker2014boundary,rachel2018interacting,mikhail2022quasiparticle}}.
%\inote{add the reference we compare our result with @Nisa}. 
Following this cue, we probe a simple model with a non-trivial topological phase structure, i.e. a bi-layer SSH \cite{ara2023topological,padavic2018topological,kurzyna2020edge,jangjan2020floquet,jangjan2022topological,li2017topological} model via studying its entanglement features for a wide range of interaction strengths.  

Over the past decade, complex systems such as quantum many-body systems have been studied rigorously from the lens of quantum information science \cite{amico2008entanglement,kitaev2006topological,levin2006detecting,pollmann2010entanglement,chen2011complete,chen2011two,schuch2011classifying}. The von Neumann entropy measures entanglement between spatial subsystems while one focuses on a particular state. For the model of our interest, one can analytically calculate the von Neumann entropy for the edge modes (see section \ref{edge_n}) that clearly show quantized behaviour for the topologically ordered phase. However, the analytic treatment via calculation of the correlation matrices is not possible beyond a very limited perturbative regime when the system becomes interacting. The computational costs for calculating the entanglement entropy for a bipartite spatial lattice are beyond the scope of classical computation for arbitrarily large system sizes. Quantum computation/simulation is expected to shed some light on this in the future. However, the current age of quantum technology, being in its infancy, has yet to demonstrate quantum supremacy in the context of a physically meaningful system size. Yet there exist state-of-the-art Hamiltonian simulation techniques, which are based on quantum information science (such as tensor network methods, which are governed by the principle of the density matrix renormalization group (DMRG)) and are implementable on classical computers 
\cite{legeza2003optimizing,dalmonte2018quantum}. In this work, apart from using analytical techniques \cite{peschel2003calculation,peschel2009reduced} for a free theory, we perform DMRG calculations and demonstrate that they are useful for studying the entanglement properties of a reasonably large system. Before working with strongly correlated systems, we perform calculations with a matrix product state ansatz for the free theory, which also matches with analytic calculations and thus benchmarks the toolbox of DMRG. 

Even for a free theory, combining the study of the entanglement entropy along with the rich symmetry properties of the physical system provides a reach sub-structure of the phenomena of bipartite entanglement given by the symmetry-resolved entanglement entropy \cite{bonsignori2019symmetry,fraenkel2020symmetry}, which is of extreme interest from both theoretical \cite{laflorencie2016quantum,goldstein2018symmetry,laflorencie2014spin,ares2022symmetry} and experimental \cite{brydges2019probing,elben2020mixed,vitale2022symmetry} perspectives. The distillation of entanglement entropy and obtaining the symmetry-resolved entanglement also lie at the core of the working principle of DMRG calculations. In this work, we demonstrate an additional feature, which is the equipartition of entanglement \cite{xavier2018equipartition} via semi-analytic study, along with the fact that the symmetry-resolved entanglement from each sector is enough to calculate the order parameter that can detect a  non-trivial topological phase. This study, presented in section \ref{sre}, is expected to be useful for scaling up computation for more complicated models/systems shortly. 

As mentioned earlier, the lack of control over analytic techniques for the model that involves a strong correlation among electrons and the unavailability of powerful numerical techniques have caused this segment to remain less explored in the literature. We demonstrate that the tensor network calculations are indeed effective in invading this particular regime and pushing the boundaries towards a very strong coupling domain. In this work, we perform DMRG calculations  \cite{white1992density,schollwock2005density} for the bi-layer SSH model to detect topological phases for a reasonably large system size and even study ground state for a length of $L=320$ for some cases. On top of that, the range of interaction strength that we could study ranges from being in a perturbative regime ($U\approx 0.2$) to being very strong ($U\approx 100$). Accessing this wide range of parameter space allows us to demonstrate the fact that the edge entanglement remains protected even under a strong correlation, which might be useful in building error-corrected topological qubits. The study of edge entanglement, as performed in section \ref{interacting}, reveals the notion of a remnant entanglement for the non-trivial topological phase due to the edge entanglement. We make an interesting observation that the edge entanglement exhibits a quantized drop from its value in the free theory as soon as a small interaction is introduced. Although manifested perturbatively, this indeed is a non-perturbative effect. As analyzed in the discussion of section \ref{interacting}, the root cause of this drop lies in the change in the degeneracy factor of the ground state just in the presence and absence of the interaction. It is robust under the change of the interaction strength. 

We study the system for the extremely large value of interaction strength more closely in section \ref{remnant}. If one neglects the hopping terms at very large coupling, the theory becomes ultra-local. Ultra-local models like these are called Carrollian field theories, which have recently gained interest in condensed matter \cite{marsot2023hall,bagchi2023magic}, fluid dynamics \cite{tong2023gauge, armas2023carrollian, bagchi2023carrollian, bagchi2023heavy}, high energy \cite{banerjee2023one, de2023carroll, ciambelli2023dynamics} and gravitational physics \cite{bagchi2023holography, bagchi2022scattering} for diverse inquiries. In particular, fermionic models with these symmetries have been dealt with in detail in \cite{bagchi2022carrollian, banerjee2023carroll, bergshoeff2023carroll}. For ultra-local systems like this, the von Neuman entropy is expected to drop to zero, which is also supported by DMRG calculations. However, if one considers including the hopping terms, albeit perturbatively small, their effects still characterize the topological phase of the system, depending upon the values of the parameters, and that is reflected in the ground state entanglement of the system as illustrated in Fig. \ref{spin}. %It is interesting to note that, even in the perturbative regime of hopping term, the remnant entanglement of edge modes remain the same, when the theory is only weakly correlated.

The organization of the paper is as follows: In section \ref{1}, we describe the model we consider in this work and summarize analytic calculational control over this model. In section \ref{2}, we probe the topological phase of the non-interacting model by studying its entanglement entropy and discuss the role of global symmetries in this study. In section \ref{interacting}, the interacting theory is studied. We clearly specify the method involved in this study, state the results and discuss its importance or consequence. In section \ref{remnant}, the case of extremely strong correlation is separately studied by focussing on the structure of its ground state in different topological phases. We finally summarize our results and state our outlook in section \ref{5}.

\section{The system:  bi-layer SSH model}\label{1}
%\subsection{AA Stacking}
%\subsection*{AB Stacking}
\begin{figure}[ht]
    \centering
    \includegraphics[width=0.45\textwidth]{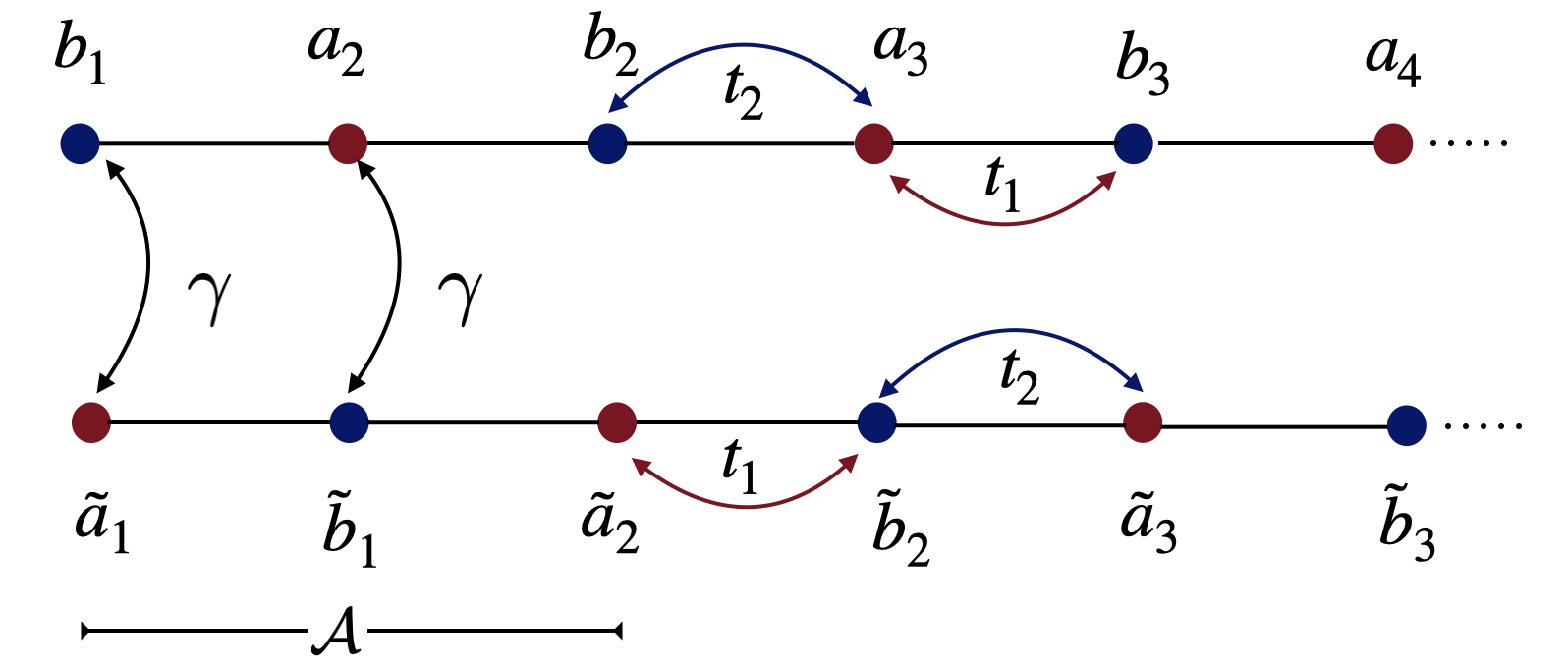}
    \caption{AB stacking model }
    %\rnote{Remove $\delta_1$ and $\delta_2$ from the figure.}
    \label{fig:enter-label}
\end{figure}
To make the article self-contained, we review the topological phases of the coupled SSH chains in this section, mostly following \cite{ara2023topological}. Our interest in the coupled SSH chains stems from the fact that they model the near-fermi-level electronic structure of a couple of polyyne chains well. The possibility of varying the physical separation between two parallelly placed such chains and thus varying the van der Waals force by strain, the band gap gains greater tunability in this model in comparison to a single chain. The topological properties of coupled polyyne chains in symmetric (AA) and asymmetric (AB) stacking arrangements are studied in \cite{ara2023topological}. According to symmetry classification, both the stacking arrangements fall in the BDI class \cite{altland1997nonstandard}.  In this work, we discuss the AB stacking arrangement in detail. This system can be driven into two distinct topological phases with tuning inter-chain hopping parameter $\gamma$ \footnote{In the realistic polyyne chain model, variation of the separation between the chains essentially amounts to tuning $\gamma$}. 

\begin{figure}[h]
\centering
    \includegraphics[width=0.8\textwidth]{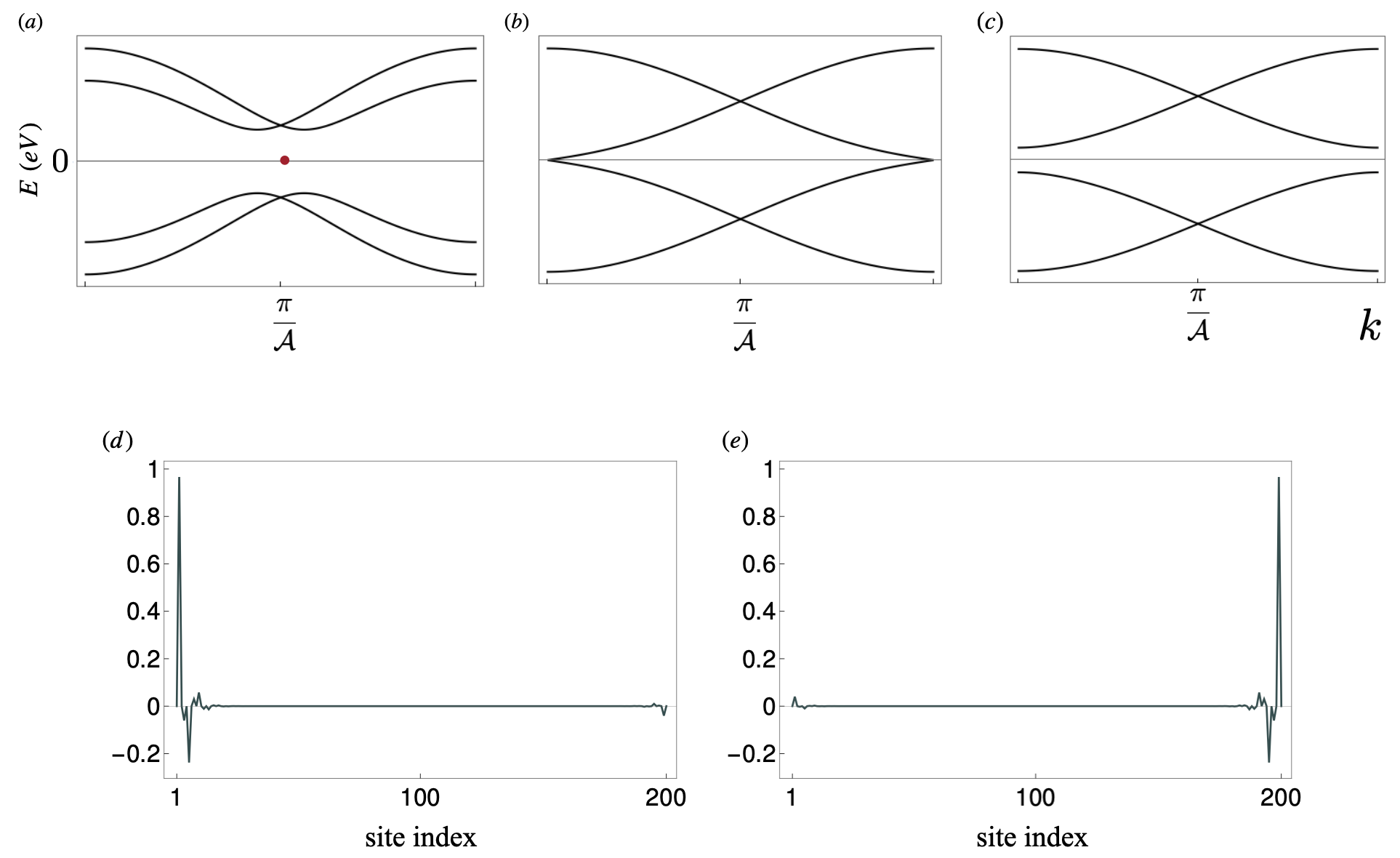}
    \caption{Band dispersion relations for $(a)~\gamma<(t_1+t_2)$; in this non-trivial insulating phase, there exists a couple of zero/gapless energy modes in the open boundary conditions; $(b)$ band gap closes at $\gamma=t_1+t_2$ and (d) reopens for $\gamma>(t_1+t_2)$. For $L=200,~ t_1=0.2,~t_2=0.8~\text{and}~ \gamma=0.05,~ (d)~\text{and}~(c)$ show localization of the edge modes. }
    \label{details}
\end{figure}

Consider the system with $L$ sites where $i=1,2 \hdots \frac{L}{2}$ belong to fermionic sites in the first chain and $i=\frac{L}{2}+1,\hdots L$ to the second chain. 
In the presence of on-site interactions, the tight-binding Hamiltonian is
\begin{align}\label{AB_h}
\mathcal{H}_{AB}  = & t_1 \sum_{i,\sigma}( a^{\dagger}_{i,\sigma} {b}_{i,\sigma}+  \tilde{a}^{\dag}_{i,\sigma} \tilde{b}_{i,\sigma})
 +t_2\sum_{i,\sigma} (a^{\dagger}_{i+1,\sigma} b_{i,\sigma}+
\tilde{a}^{\dag}_{i+1,\sigma}  \tilde{b}_{i,\sigma})
+\\ \nonumber
& \gamma \sum_{i,\sigma}( a^{\dag}_{i+1,\sigma} \tilde{b}_{i,\sigma}+\tilde{a}^{\dag}_{i,\sigma} b_{i,\sigma})
+ \mathrm{H. c.}+ U \sum_i  n_{i, \uparrow} n_{i, \downarrow}
\end{align}
where $n_i= x^\dagger_{i}x_{i}$, ($x=(a,b,\tilde{a},\tilde{b}), ~ \sigma=\uparrow \downarrow$) and $U$ is the Hubbard interaction. \cite{altland1997nonstandard,kitaev2006topological}. For $U=0$, the system is a non-interacting coupled SSH model. While discussing the free theory, we will suppress the spin index and only work with a single spin species. For periodic boundary condition (PBC), \eqref{AB_h} can be written in the Fourier space as:
%\rnote{1. Eq. \eqref{AB_hk} is not the Hamiltonian, but the kernel. 2. Where is $\gamma_3$ coming from?. 3. Eq. \eqref{AB_hk} is consistent with \eqref{AB_h} only when $U=0$ and you are not considering spin. Make this clear.},
\begin{equation}
\mathcal{H}_{AB}=  \sum_k {\Psi_k^{\dag}}
 \begin{pmatrix}
    0 & f_k & 0 & \gamma g_k \\
    f^{*}_k & 0 & \gamma & 0 \\
    0 & \gamma & 0 & f_k \\
    \gamma{g_k}^{*} & 0 & f^{*}_k & 0 \\
 \end{pmatrix}
\Psi_k.
\label{AB_hk}
\end{equation}

where $\Psi^{\dag}_k=( c^{\dag}_k \hspace{0.4cm} d^{\dag}_k \hspace{0.4cm} \tilde{c}^{\dag}_k \hspace{0.4cm} \tilde{d}^{\dag}_k )$ is a four-component fermion. Here $c_k, d_k, \tilde{c}_k$ and $ \tilde{d}_k$ are the Fourier basis modes corresponding to the real space modes $a_i, b_i, \Tilde{a}_i$ and $\Tilde{b}_i$ respectively, eg. $c_k = \frac{1}{\sqrt{N}} \sum_{p=1}^N e^{-i k\mathcal{A}j} a_j$ etc. and $f_k=t_1+t_2 e^{-i \mathcal{A} k}~~ \text{and}~~ g_k=e^{-i\mathcal{A}k}$. Here $\mathcal{A}$ is the lattice spacing. The four energy bands emanating from diagonalizing the Hamiltonian kernel \eqref{AB_hk} are given by,   
%\rnote{Remove the $\iota$ thing. It's not acceptable to use $\iota$ and $i$ interchangeably}.
\begin{equation}
    \mathcal{E}_{AB}= \pm \sqrt{{|f_k|}^2+{\gamma}^2\pm 2\gamma(t_1+t_2)\cos\left(\frac{\mathcal{A}k}{2}\right)}.
\end{equation}
Hence, in terms of the fourier space oscillator basis $\tilde{c}_k,  \tilde{d}_k$ that diagonalizes the kernel \eqref{AB_hk}, the ground state of the system is given as
\begin{equation}
    \ket{\Psi}= \prod_{k} \tilde{c}^{\dag}_k  \tilde{d}^{\dag}_k\ket{0}.
\end{equation}
The critical point $\gamma=t_1+t_2$ distinguishes between the trivial and non-trivial topological phases. The Zak phase is quantised as $2\pi$ and $3\pi$ for $\gamma<t_1+t_2$ and $\gamma>t_1+t_2$ respectively. The band dispersions for the different phases have been shown in the first three panels of Fig. \ref{details}.

According to the bulk-edge correspondence, in open boundary condition (OBC), a pair of zero energy edge modes exist in the non-trivial topological phase $\gamma<t_1+t_2$, which have been shown as the dots on the fermi-level in panel (a) of Fig. \ref{details}. The fact that these zero modes are localized at the edges of the chains can be seen in the following way.

Let's start with a generic Hamiltonian similar in form to \eqref{AB_h}, with $U=0$, ie. $ \mathcal{H}= \sum_{i,j=1}^{L} x_i^{\dagger} H_{ij}x_j $,
here $H$ is an ($L\times L$) matrix. Diagonalizing this, we have $H = VDV^\dagger $ where
$V$ is unitary, and $D$ is a diagonal matrix. In the new basis, $\mathcal{H}= \sum_{l=1}^{L} \epsilon_l \alpha_l^{\dagger} \alpha_l $. Here $\alpha^{\dagger}$ are the modes which create the single-particle energy eigenstates: $\alpha_l^{\dagger}=\sum_{i=1}^{L}  V_{il} x_i^{\dagger}.$ If the set of $\epsilon_l$ describes a set of gapped dispersion bands and there are a few points not belonging to any of the continuous bands, then the later ones show up as a curious feature in terms of their corresponding eigenvectors. The eigenvectors $\{V_{il}| i = 1, \dots, L\}$ corresponding to the modes $\alpha_l$, whose energy eigenvalues lie on any of the continuous bands, show standing wave behaviour, as expected by identifying the eigenvectors to the wavefunctions of particles in a box (in first quantization formalism). On the other hand, the eigenvectors corresponding to the isolated modes show a typical feature of getting localized towards the edges of the chain. This has been visualized explicitly in panels (b) and (d) of Fig. \ref{details}. In this figure, we particularly used the zero energy isolated modes of the coupled SSH chains to establish that they are indeed edge modes. The analytic proof of the same can be found in \cite{ara2023topological}.
%For the isolated modes, only edge modes contribute to the single-particle energy eigenstate. This is shown in Fig.~\ref{details}$(d)~ \text{and} ~ (e)$, we plot $\{ V_{1l}, V_{2l} \hdots V_{Ll}\} $, and it has a finite value at the edges only. 
%\rnote{Expression of ground state is wrong! Correct it.}
%\rnote{@Nisa: Write the analytical expressions of the dispersion bands here. Copy from our JCMP paper. Also write the relation between the k-space modes and the $a,b, \Tilde{a},\Tilde{b}$ ones here. We need these to write the ground state.}

\begin{comment}
\subsection{AA Stacking}
We consider spin-less, non-interacting fermions described by the following tight-binding Hamiltonian 
\begin{align}
    \mathcal{H}_{AA}= &t_1 \sum_{i}( a^{\dagger}_p {b}_i+  \tilde{a}^{\dag}_i  \tilde{b}_i)+
      t_2\sum_{i} (a^{\dagger}_{i+1} b_i \\ \nonumber
  & +\tilde{a}^{\dag}_{i+1}  \tilde{b}_i)+ \gamma_1 \sum_{i} (\tilde{a}^{\dag}_{i} a_i+\tilde{b}^{\dag}_{i} b_i)+ \mathrm{H. c.} ~.
    \label{AA_h}
\end{align}
Four finite-energy modes exist for $\gamma_1<\frac{|t_1-t_2|}{2}$.
\end{comment}

\section{Entanglement Entropy: The free theory}\label{2}

\subsection{The von Neumann Entropy}
\begin{figure}[ht]
    \centering
    \includegraphics[width=0.45\textwidth]{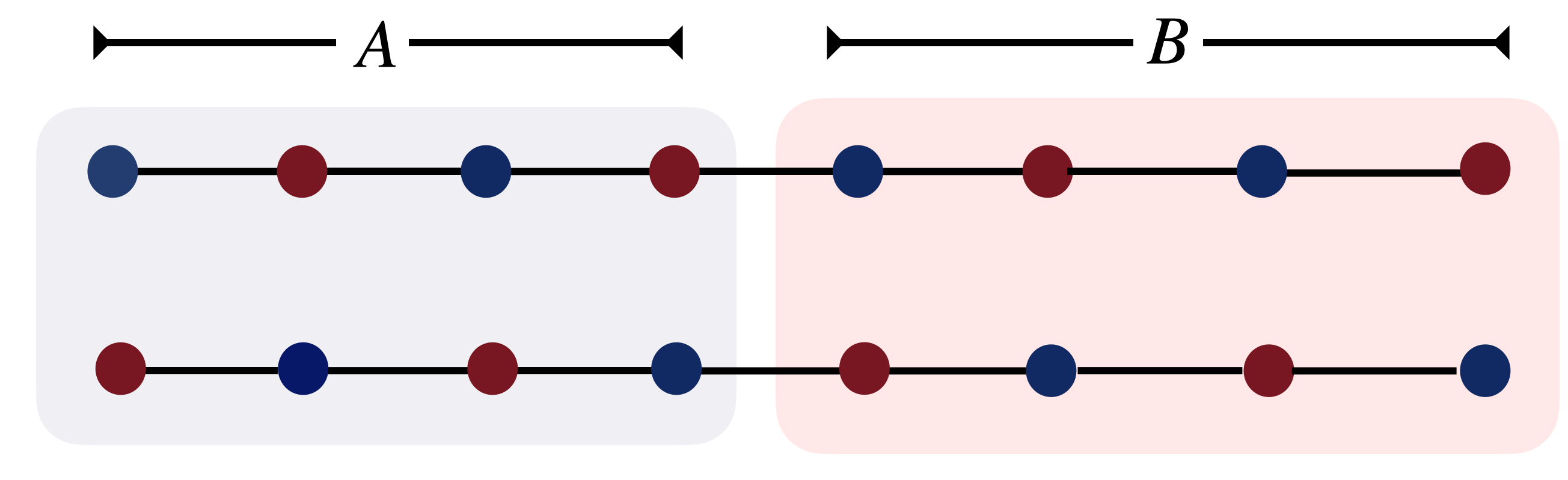}
    \caption{Bipartite System with equal subsystems $A$ and $B$.}
    \label{bipartite}
    \end{figure}
For a many-body system, the von Neumann subregion entanglement entropy can be defined as follows. At a given instant of time, let us divide the system into two regions: $A$ and $B$. At zero temperature, if the system is at a state  $|\Psi\rangle$, the von Neumann entanglement entropy of region $A$ with respect to $B$ is defined as, 
\begin{eqnarray} \label{vn_def}
  S_A=-\mbox{tr}_A\rho_A \ln{\rho_A},  
\end{eqnarray}
 where $\rho_A =\mbox{tr}_B|\Psi\rangle \langle\Psi|$  is the reduced density matrix. In this article, we will choose $|\Psi \rangle$ as the ground state with respect to the given Hamiltonian describing the system.

%The same reduced density matrix could as well be written as $$ \rho_A=\frac{e^{-H_A}}{Z},$$ where $H_A$ is defined as the entanglement Hamiltonian and $ Z= \text{tr}[e^{-H_A}]$ \cite{peschel2003calculation}.
For a free fermionic many-body system, whose Hamiltonian is of the form $ H=\sum_{ij} \mathcal{H}_{ij} c_i^{\dagger} c_j,$, the entire information of ground state entanglement is encoded in the equal-time correlation functions, ie. $ C_{ij} = \bra{\Psi}c_i^\dagger c_j\ket{\Psi}$ \cite{peschel2003calculation}. Here $c_i$ are the local fermionic degrees of freedom. If the region $A$ contains the set $\{c_i |i \in A \}$ of degrees of freedom, then the von Neumann entropy \eqref{vn_def} can be expressed as:\begin{equation}
         S_A=-\sum_k \zeta_k\ln{\zeta_k}+(1-\zeta_k)\ln{(1-\zeta_k)}.
         \label{ent}
\end{equation}
Here $\zeta_k$ are the eigenvalues of the matrix $C_{ij}$ with $i,j \in A$. It is to be noted that, for a fermionic system, $0 \leq\zeta_k \leq 1$. The eigenvalues $\zeta_k$, for a system with particle-hole symmetry, can be calculated using the knowledge of the eigenvectors of the Hamiltonian kernel itself \cite{sirker2014boundary,fishman2015compression}. If $V_i$ are the eigenvectors of the Hamiltonian kernel $\mathcal{H}$, the matrix for the correlation function $C_{ij}$ for particle-hole symmetric systems can be expressed as:\begin{eqnarray}
    C_{ij} = \sum_{k =1}^{L/2} V^{\star}_{ik} V_{kj},
\end{eqnarray}
where the sum runs over half the system size $L$, corresponding to the hole modes, i.e. the ones with negative eigenvalues of $\mathcal{H}$. We use this result for direct computation of \eqref{ent} above.

For our present work involving the coupled SSH chains, we choose the bi-partition as shown in Fig. \ref{bipartite} with the regions $A$ and $B$ having an equal number of degrees of freedom.

%While considering a one-dimensional tight-binding Hamiltonian $ H=\sum_{ij} \mathcal{H}_{ij} c_i^{\dagger} c_j,$ for which the correlation matrix is given by $ C_{ij} = \bra{\Psi}c_i^\dagger c_j\ket{\Psi}$, one can calculate the von Neumann entropy semi-analytically. The eigenvalues of Correlation matrix $\zeta$ are related to the eigenvalues of entanglement Hamiltonian $\epsilon_k$ by $\zeta_k=\frac{1}{1+e^{\epsilon_k}}$. The entanglement entropy in this framework is given by,
%The von Neumann entropy of the system such that $S_A=S_B$ is defined as, $S_A=-\mathrm{tr}_A\rho_A \ln{\rho_A}$ where $\rho_A =\mbox{tr}_B|\Psi\rangle \langle\Psi|$.
%\begin{equation}
%         S_A=-\sum_k \zeta_k\ln{\zeta_k}+(1-\zeta_k)\ln{(1-\zeta_k)}.
         \label{ent}
%\end{equation}
%Note that, using \eqref{ent}, one can calculate entanglement entropy for a state of the free theory upto a reasonably large system size. 

%\rnote{The analysis up to eq. (5) is true for free theory and one doesn't need DMRG for this.}

%\subsection{The model with strongly interacting fermions}
%We model the spin-full interacting SSH model defined in equation (\ref{AB_h}) using DMRG. 

\begin{comment}
    The entanglement entropy and energy have the following form
\begin{equation}
    S_A= \sum_a(\ln{(1+e^{-\epsilon_a})}+\frac{\epsilon_a}{1+e^{\epsilon_a}})
\end{equation}

The correlation matrix is given by
\begin{equation}
    C_{ij}=\langle a^{\dagger}_ib_j\rangle=\mbox{tr}(\rho_A a^{\dagger}_ib_j)
\end{equation}
The matrix elements of the correlation matrix $C_{ij}$ \cite{sirker2014boundary}(eq. 4.17), can be expressed as 
\begin{equation}
    C_{ij}=\sum_{k,occupied} (\psi^k_i)^* \psi^k_j
\end{equation}
Diagonalizing the correlation
matrix then gives the single-particle eigen energies $\epsilon$ of the entanglement
Hamiltonian using the relation 
\begin{equation}
    \zeta=(1+e^{\epsilon})^{-1} 
\end{equation}

Eigen values of the correlation matrix  are $\zeta$ which are between 0 and 1, and entropy is given as 
\begin{equation}
S_A= -\sum_a (\zeta_a \ln{\zeta_a}+(1-\zeta_a)\ln({1-\zeta_a}))
\end{equation}
\end{comment}
%\section{Entanglement as Order Parameters}\label{2}

\subsection{Entanglement entropy for the edge modes}\label{edge_n}
In this sub-section, we demonstrate that the von Neumann entropy defined earlier can be used to probe different topological phases of the theory. As a characteristic feature of this particular class of theories, existence of the edge modes is directly related to the system being in the non-trivial topological phase or not. In this regard, extracting the entanglement of the edge modes stands to be crucial \cite{ryu2006entanglement}. 

\begin{figure*}[ht]
    \centering
    \includegraphics[width=0.55\textwidth]{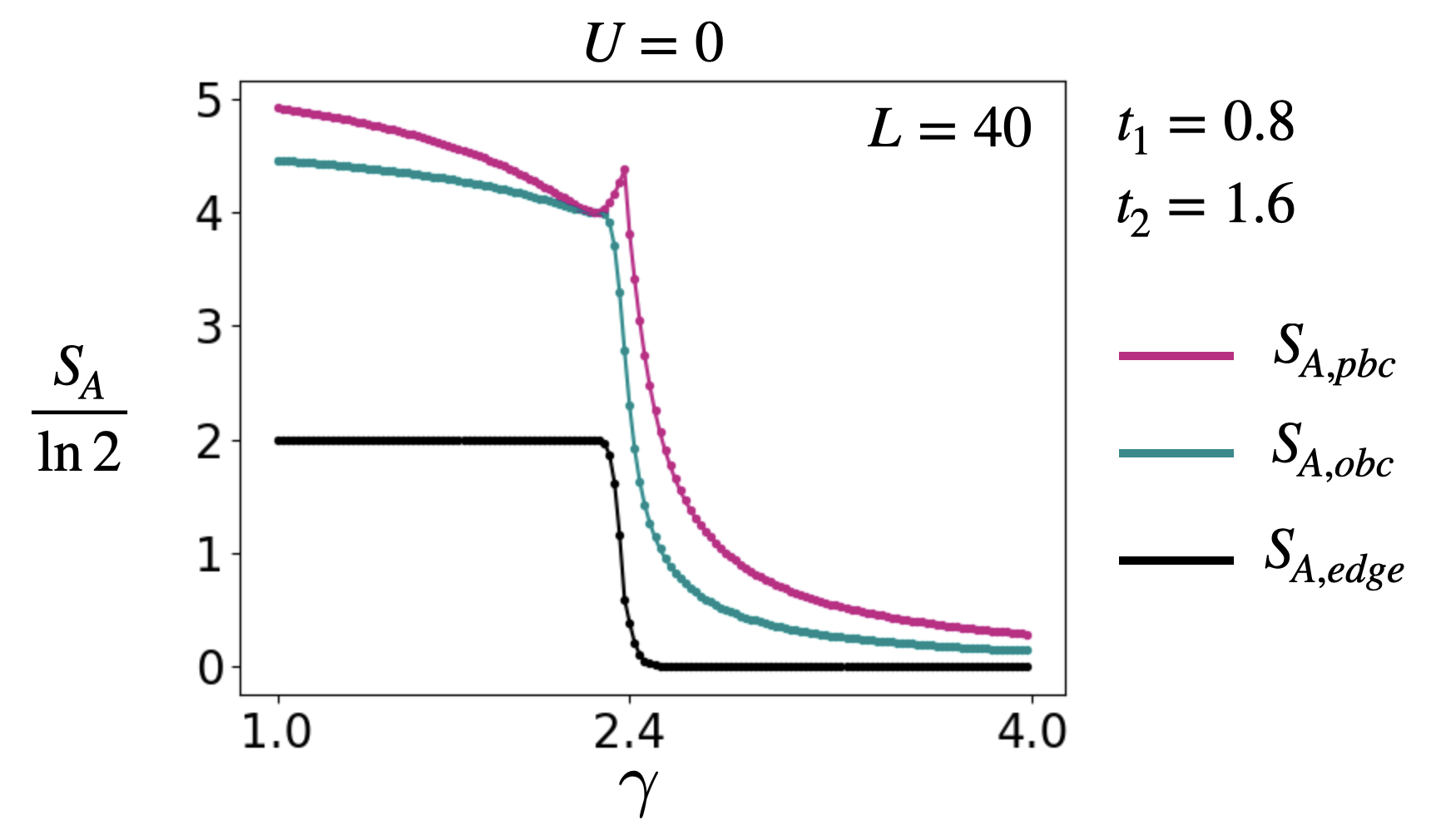}
    \caption{The von Neumann entropy for non-interacting fermions is calculated using \eqref{ent}. $S_{obc},~S
_{pbc},~\mathrm{and}~S_{edge}$ are plotted for $U=0$ as a function of inter-chain hopping strength $\gamma$. }
    %\inote{extend the caption to include definition of $\gamma$, ref to equations that define edge entanglement etc. } }
    \label{map}
\end{figure*}

It is important to note that edge modes can only exist for a system with open boundary conditions, but never with periodic boundary conditions. Following \cite{wang2015detecting}, one may calculate entanglement entropy for a system with both open and periodic boundary conditions separately, and define the edge entanglement as:
\begin{equation} \label{edge_EE}
    S_{edge}=S_{A,obc}-\frac{S_{A,pbc}}{2}
\end{equation}
In this work, we consider the spin-full free theory for a ladder system bi-partitioned into two halves, as shown in Fig.~\ref{bipartite}.  Calculation of von Neumann entropy $S_A$ for open and periodic boundary conditions is possible following the method outlined in the previous section. Entanglement of the edge modes, defined in (\ref{edge_EE}) is also calculated, and one can find a distinct notion of two different topological phases as shown in the plot in Fig. \ref{map}. Note that, the same plot has also been reproduced using another method, the tensor network calculations, which is based on DMRG algorithm.  We elaborate on the DMRG method in the next section, where it is crucial as the correlation matrix method is not generalizable for a theory with onsite interaction. Note that, our calculation is consistent with \cite{wang2015detecting}, where edge entanglement is calculated for the single-chain SSH model.

As obtained in Fig. \ref{map}, edge entanglement indeed acts as an order parameter that takes a quantized value for the topologically non-trivial phase and sharply drops to zero if the system is in the other phase. 
\begin{equation} \label{edge_EE-free-result}
    S_{edge}=  \begin{cases}
         2\ln{2}, &  \gamma <(t_1 + t_2)  \\
   0, &  \gamma > (t_1+t_2).\\
  \end{cases}
\end{equation}
Note that another notion of edge entanglement lies in the fact that the von Neumann entropy, calculated for a system, always possesses a lower bound with an increasing value of the hopping parameter. This lower bound is zero for a topologically trivial phase, while it is $2\ln 2$ for the system to be in the topologically ordered phase. This denotes the lower bound to be directly connected to the edge entanglement. 
These are the features of entanglement entropy that characterize different topological phases of the system, which deserves some investigation for the model with on-site Hubbard interaction that models realistic systems more closely.

Before exploring the interacting theory, we make a few more observations relevant for scaling up the calculation in terms of system size and richer symmetry structure.
\begin{enumerate}
    \item Reperforming the calculation with DMRG based tensor network algorithm reproduces all of the results reliably and thus benchmarks the tensor network codes.
    \item Going to a larger system size (as permitted by the computational resources available, both with correlation matrix method and DMRG), the features of the entanglement entropy remain the same. This is demonstrated in appendix \ref{a1}. 
    %\inote{@Nisa: Put L=56 plots here, if possible, a single plot combining 40 and 56}
    \item A further study with consideration of the symmetry resolution of entanglement entropy for each global symmetry sector is performed. The symmetry-resolved entanglement, reported in the next subsection, demonstrates that the contribution of edge entanglement, when calculated only from a particular global symmetry sector, can clearly show the signature of different topological phases. Hence, for a system with a rich symmetry structure, studying a particular symmetry sector, which might be of a lot lesser dimension than the entire Hilbert space, can be useful in identifying different topological phases. This is expected to be useful when global symmetries are non-Abelian. 
\end{enumerate}

%\subsubsection{Remnant entanglement}
\subsection{Symmetry resolved entanglement}\label{sre}
Recent studies revealed \cite{laflorencie2016quantum,goldstein2018symmetry,laflorencie2014spin,ares2022symmetry}, rich substructure in entanglement entropy of ground states of systems having continuous global symmetries. Motivated by experimental evidence \cite{brydges2019probing,elben2020mixed,vitale2022symmetry}, one can formulate a pertinent question: \textit{how much does each super-selection sector of the full Hilbert space, each designated by global quantum numbers, contribute to the total entanglement?} In fact, for abelian symmetry, it turns out that only a few super-selection sectors (those having nearly the maximum value of the quantum number) have equal finite contributions. This result is the equipartition of entanglement and can be well demonstrated analytically for systems near critical point \cite{northe2023entanglement}.

The free coupled chain system \eqref{AB_h}, with a single spin species, has the $U(1)$ charge conservation symmetry. The inclusion of onsite interaction and different spin species, of course, enhances it to a non-Abelian $SU(2) \times SU(2)$ one, i.e. the symmetry group of the Hubbard model. For the present work, we restrict ourselves to the free case. The analysis of this section stems from the argument that the $U(1)$ conserved charge $Q$ commutes with the density matrix of the eigenstates of the Hamiltonian. For example, the ground state for the free theory is half-filled and hence is an eigenstate of $Q$, resulting in $[Q, \rho] = 0$. If the system is partitioned into $A$ and $B$, the charges for each subsystem $Q_A, Q_B$ are individually conserved and add up to the total charge $Q = Q_A+Q_B$. Tracing over the sub-system $B$ Hilbert space, we see that the reduced density matrix $\rho_A$ commutes with charge for subsystem $A$: $[Q_A, \rho_A] = 0$. This helps to arrange the reduced density matrix in a block diagonal structure, each block/ super-selection sector having a definite $Q_A$ charge:
\begin{eqnarray}\label{rho_decomp}
    \rho_A = \oplus_q p(q) \rho_A (q),
\end{eqnarray}
where the sum is over the spectrum of $Q_A$ and $\rho_A(q)$ is basically $\rho_A$ projected onto the $Q_A =q$ subspace. By definition of projection operation, $\sum_q p(q) = 1$ and $p(q)$ hence can be interpreted as the corresponding probability. From, \eqref{rho_decomp}, one observes that the von-Neumann entropy for the subsystem $A$ gets decomposed as follows:
\begin{eqnarray} \label{conf_dist}
    S_A %{\mathrm{vN}}% &=& -\mathrm{tr} \left(\rho_A \ln \rho_A \right) \nonumber\\
    &=& \sum_q \left(p(q) S_A(q) - p(q) \ln p(q)\right), 
\end{eqnarray}
where $S_A(q) = - \mathrm{tr} \left( \rho_A (q) \ln \rho_A (q)\right)$ is the von-Neumann entropy for the charge sector $q$.
For $U(1)$ symmetric systems \cite{goldstein2018symmetry},  the symmetry resolved contribution to the von-Neumann entanglement entropy coming from the super-selection sector of charge $q$ is given by:
\begin{eqnarray} \label{sym_res}
    S_A(q) = -\partial_n (\mathcal{Z}_n(q)/\mathcal{Z}_1^n(q)) |_{n=1},
\end{eqnarray}
where $\mathcal{Z}_n(q)$ is given by the $n$-sheets replica copy of the density matrix trace projected on the $q$-charge sector:
\begin{eqnarray} \label{znq}
    \mathcal{Z}_n(q) = \frac{1}{2\pi} \int_{0}^{2\pi} d\alpha \, e^{-iq \alpha} \mathrm{tr} (\rho^n_A e^{i Q_A \alpha}).
\end{eqnarray}
%Here $\rho^n_A$ is the reduced density matrix for the subsystem $A$, and $Q_A$ is the total charge operator for that subsystem. 
Interestingly, The probability $p(q)$ is given by $\mathcal{Z}_1(q)$.

In fact, for free fermions, just as $\mathrm{tr} \rho^n_A$ is given in terms of the eigenvalues of the correlation matrix, the symmetry sector projected version of it also can be expressed in terms of those eigenvalues:
\begin{eqnarray} \label{rhoeigen}
    \mathrm{tr} (\rho^n_A e^{i Q_A \alpha})  = \prod_k \left(\zeta_k^n e^{i \alpha} + (1- \zeta_k)^n \right).
\end{eqnarray}
If one is interested in the first term of the sum \eqref{conf_dist}, i.e. the total configuration entropy $\sum_q p(q) S_A(q)$, the easiest way would be to calculate the individual probabilities $p(q)$ and subtract the distillable entropy part $- \sum_q p(q) \ln p(q)$ from the full von-Neumann entropy. %\rnote{I will insert plots for the symmetry resolved and distilled entropies here.}
\begin{figure}[ht]
    \centering
    \includegraphics[width=0.9\columnwidth]{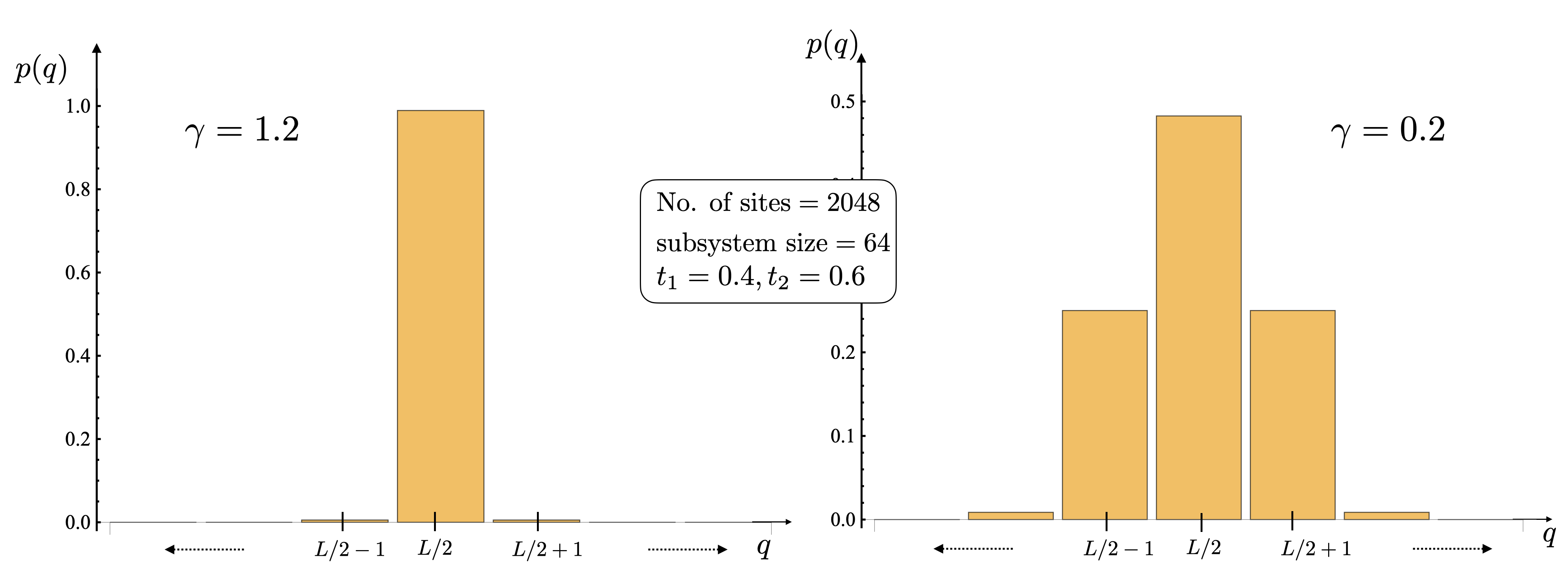}
    \caption{Probability distribution of each super-selection sector designated by the global quantum number $q$}
    \label{Prob dist}
    \end{figure}

For notational convenience, we will omit the subsystem suffix $A$ and simply denote the $q$-charge sector von-Neumann entanglement entropy \eqref{sym_res} by $S(q)$. Using the definitions \eqref{sym_res}, \eqref{znq} and the result \eqref{rhoeigen}, one can readily calculate the symmetry resolved entanglement entropy $S(q)$ for free fermionic systems. Since, for a limited number of values of the charge $q$, $S(q)$ are non-zero and are equipartitioned over those charge sectors, most of the global features regarding quantum phases and edge entanglement are contained in just one of such non-zero $S(q)$. 

\begin{figure}[ht]
    \centering
    \includegraphics[width=0.6\textwidth]{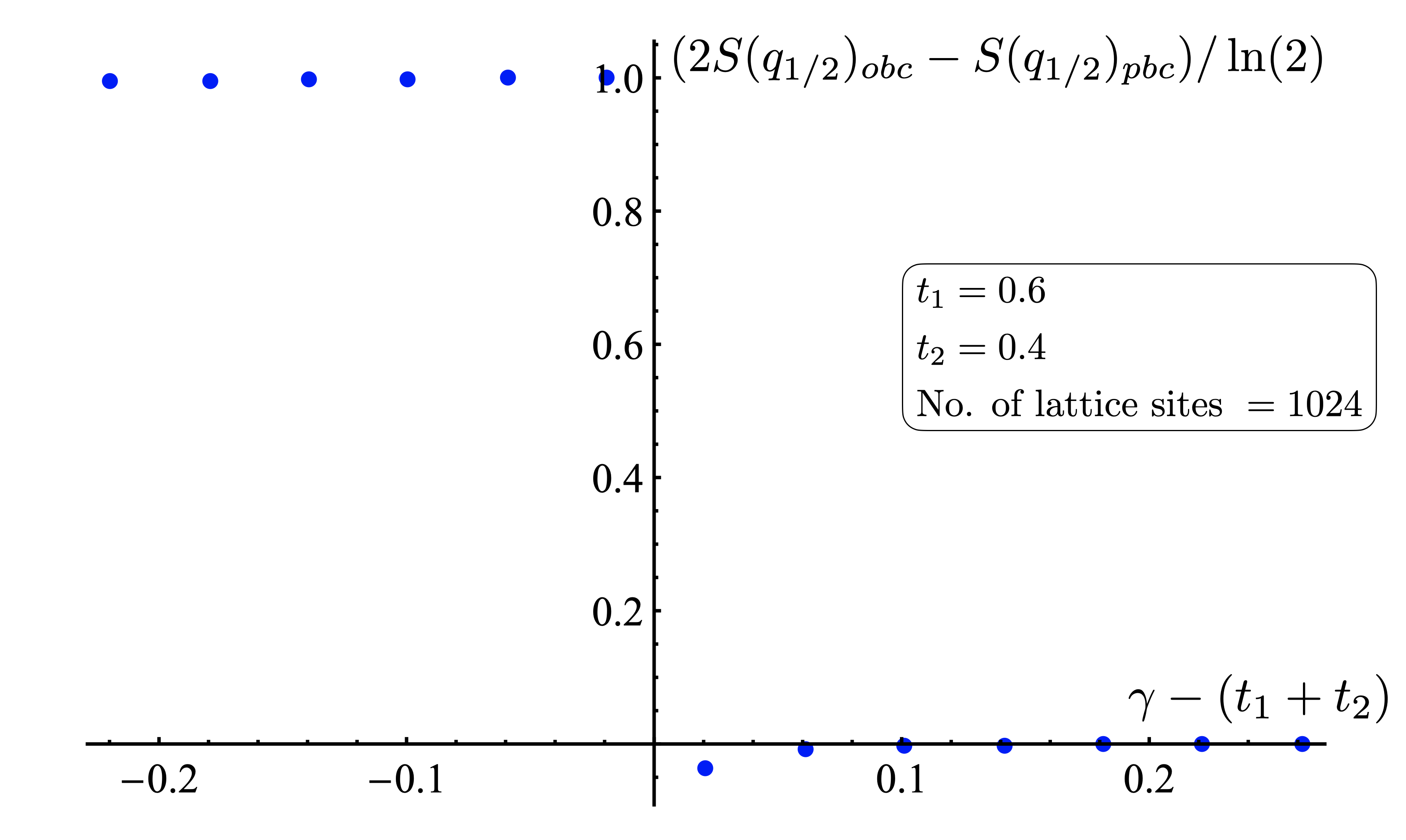}
    \caption{Twice the symmetry resolved edge entanglement entropy $2S(q_{1/2})_{obc} - S(q_{1/2})_{pbc}$, for AB-stacking, for a subsystem cut as shown in figure  Fig. \ref{map}. It shows a clear signature of phase transition, even just for a single charge sector.}
    \label{sym_res_s}
    \end{figure}

For our model, ie the AB-stacked coupled free SSH chains, we compute $S(q_{1/2})$ using the definition \eqref{sym_res}. Here, $q_{1/2}$ is half the maximum $U(1)$ charge possible in a subsystem $A$, as shown in Fig. \ref{bipartite}. For concreteness, we took a $1024$ site system, keeping the intra-chain hopping parameters fixed at $t_1 = 0.6, t_2 = 0.4$ and varying the inter-chain hopping parameter such that $\gamma - (t_1+t_2)$ changes sign. As usual, contributions to entanglement from the edge are captured by the scaled difference between the open boundary condition and the periodic one (cf. \eqref{edge_EE}). Interestingly, the symmetry-resolved entanglement, too, shows a sharp phase transition, as seen in the plot \ref{sym_res_s}.
When compared to the result for the full von Neumann entropy \eqref{edge_EE}, the main difference is that this symmetry resolved edge entanglement in the $\gamma < t_1 +t_2$ phase is $1/4$ to that of the full von Neumann entropy. For a similar set of parameters, we checked for the probability distribution function $p(q)$ for $q$, taking values near $L/2$. As shown in Fig. \ref{Prob dist}, in the phase without edge modes, the ground state has almost $p(L/2) \approx 1$, ie. clearly showing the half-filled nature. In the other phase, however, the ground state has finite probabilities for one-less or one-more than half-filling.

\section{The Interacting theory}\label{interacting}
In this section, we study bipartite entanglement between the edges of the coupled chain model, now in the presence of interactions. Small on-site interactions could be studied with mean-field approximations and quantum Monte Carlo (QMC) simulations \cite{troyer2005computational,wang2015detecting}. However, we use DMRG-based numerical tools for the present work. This, as motivated in the introduction, is tailor-made for a two-pronged advancement. The first one is the capability of DMRG techniques to handle a larger number of sites, as far as an accurate prediction of the ground state is concerned. This is necessary because of benchmarking outcomes from classical computing architecture against near-term scaled-up quantum computing ones. As we will see in the next section, interesting features of the low energy sector of the Hilbert space start showing up as one goes to an extremely strong coupling regime, where the notion of quasi-particle excitations no longer exists. The second advancement made by DMRG for these particular investigations is difficult to reach by other numerical techniques, to the best of our understanding.

In this present set of studies, computational resources limit us to choose a system size of $L=40$ (computations for $L=56$ have also been carried out, showing essentially the same quantitative features and have been presented in \ref{a1}). We have used the ITensor package \cite{itensor,itensor-r0.3} to construct the MPS ansatz and used the DMRG algorithm to find the ground states for the interacting theory.
%\subsection{Protection of edge entanglement against strong correlation} 
\subsection{Sudden reduction of edge entanglement}
\subsubsection{Method}

In this section, we provide the details of the numerical techniques used to obtain the data presented in the following sections. Since we are dealing with a ladder system, we mapped it to a pseudo-one-dimensional lattice, where the odd sites correspond to the upper rung of the ladder and the even site is the lower rung. From there, we employ a Gaussain-MPS (GMPS) ansatz described in \cite{fishman2015compression} and the open-source ITensorGaussianMPS \cite{itensor} package to construct the ground-state for the ladder system in $U=0$ regime. This was done for both the OBC and PBC periodic conditions. On the PBC side, one must go to very large bond dimensions to achieve modest convergence in the ground-state energies. We check for the convergence of ground-state energies and the bipartite von Neumann entanglement entropy by comparing it with an exact diagonalization code developed in-house. We explored bond dimensions up to $\chi = 3000$ to ensure a $10^{-8}$ convergence for the ground-state energy and entanglement entropy. We then use this ground state obtained from the GMPS ansatz as an initial guess wavefunction for the running DRMG for interacting theory Hamiltonian. Once the ground state for the interacting case has converged for a tolerance of $10^{-8}$, one can extract the entanglement entropy, both for the OBC and PBC cases.
\subsubsection{Results}
\begin{figure}[h]
    \centering
    \includegraphics[width=0.8\columnwidth]{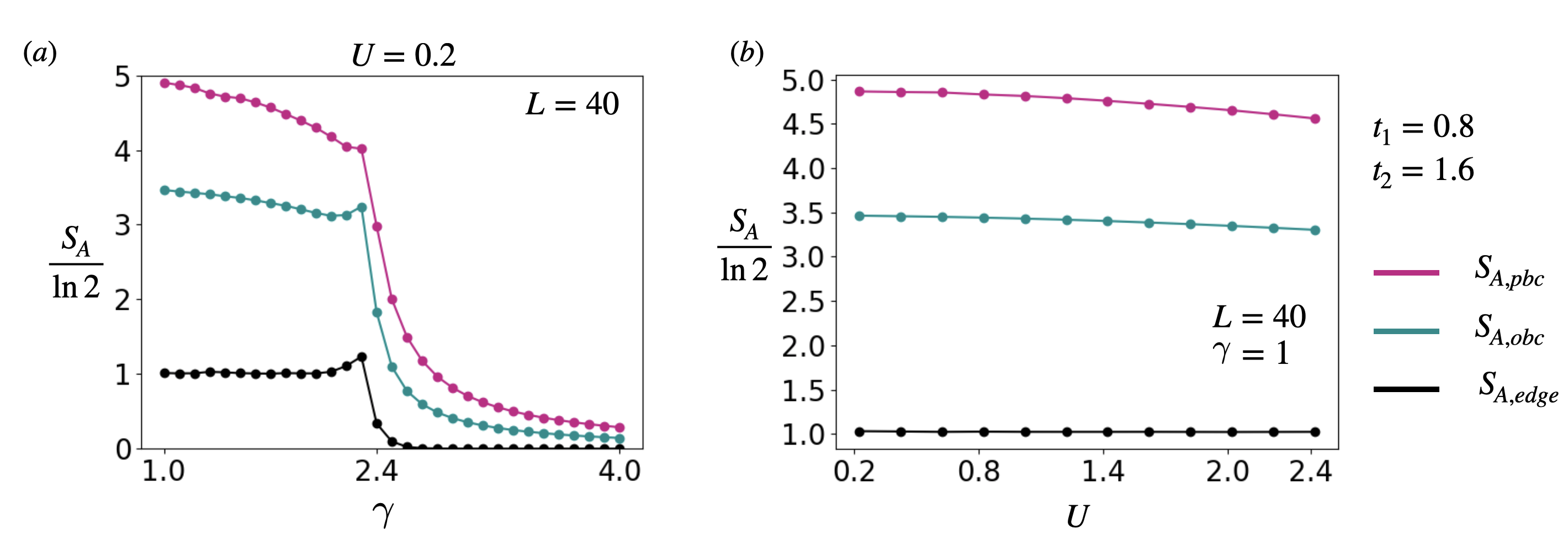}
    \caption{$(a)$ Here $S_{edge}$ reduces to $\ln{2}$ from $2\ln{2}$ with $U\neq 0$. $(b)~S_{edge}$ is plotted for $U$ in the non-trivial topological phase.  }
    \label{int}
\end{figure}

%\textit{Interacting} ($U\neq 0$): 
We study the effect of Hubbard interaction on the zero energy edge modes. The system is bi-partitioned, the same way as discussed in section \ref{edge_n}, i.e. as per the Fig. \ref{map}. As we introduce $U\neq 0$ in the system, $S_{A,obc}$ is reduced compared to the non-interacting case as shown in Fig.~\ref{int}$(a)$. A very clear transition is seen at the critical point $\gamma=(t_1+t_2)$, indicating the critical point to separate two distinct quantum phases. Similarly we calculate $S_{A,pbc}$ and compute $S_{edge}=S_{A,obc} - \frac{1}{2}S_{A,pbc}$ for $U=0.2$. In Fig.~\ref{int}$(a)$, $S_{edge}$ is quantised as $\ln{2}$ in the non-trivial topological phase and sharply goes to 0 in the other phase. Again, this result is consistent with \cite{wang2015detecting}, where $S_{edge}$ for $U\neq 0$ is calculated for the SSH model using quantum Monte Carlo (QMC) simulations. We also plot $S_{edge}$ for the varying interaction strength in the non-trivial topological phase, which remains quantised at $\ln{2}$ as shown in Fig~\ref{int}$(b)$.
\begin{equation} \label{int_edge_EE}
    S_{edge}=\begin{cases}
         \ln{2}, &  \gamma <(t_1 + t_2)  \\
   0, &  \gamma > (t_1+t_2).\\
  \end{cases}
\end{equation}
%Subsystem entanglement entropy $S_A$ does not change with the size of the system ($L$) away from the critical point. We have shown this in Fig.~\ref{int} $(b$ and $(d)$ with different system sizes $L=56$ and $L=40$ for OBC and PBC. %(\textcolor{violet}{Mention pbc limitations for a large number of sites}). We conclude that $S_{edge}$ is quantised as $\ln{2}$ for large $L$.
Edge entanglement entropy is reduced to $\ln{2}$ from $2\ln 2$. This sudden reduction of edge entanglement entropy from $2 \ln 2$ for the free spin-full theory \eqref{edge_EE-free-result} to $\ln 2$ for the interacting theory, even for small but finite $U$ is definitely not a perturbative effect, as $S_{edge}$ is not decreasing slowly. This is rooted in the symmetries of the Hamiltonian. We explain this feature in the following subsection.
\subsubsection{Discussion}
We denote the zero-energy edge modes for each spin as $\alpha_{p,\sigma}$, where  $p =1,2$ stands for edge modes for the left and the right edge, respectively. Any state of the form 
\begin{eqnarray}
|\mathrm{edge} \rangle = \sum_{p,q} A_{p,q} \alpha^{\dagger}_{p,\uparrow}\alpha^{\dagger}_{q,\downarrow} |0\rangle \label{degen}
\end{eqnarray}
has zero energy with respect to the free theory Hamiltonian in the non-trivial topological phase. The $A$ coefficients appearing in \eqref{degen} are complex numbers subject to the condition that the state is normalized. Hence, the edge-state is 4-fold degenerate. For the same reason, the ground state,
\begin{eqnarray}
    |\mathrm{ground} \rangle = |\mathrm{edge} \rangle \otimes \prod_{k} b^{\dagger}_k |0 \rangle
\end{eqnarray} is also degenerate. Here, $b_{k}$ are all the finite, negative energy modes (holes). The 4-bilinears $\alpha^{\dagger}_{i,\uparrow}\alpha^{\dagger}_{j,\downarrow} $ are bosonic in nature. Hence, the degenerate edge state space is an orbit of $SU(4)$ action on linear combinations of 4 bosons.

At the thermodynamic limit, $\alpha_i$ are exactly localized at the edges, and the maximally entangled edge state is the one when all the 4 $A$'s appearing in \eqref{degen} are equal, modulo $U(1)$ phases. In that case, the edge state is a couple of Bell pairs. This gives an entanglement entropy of $2 \ln 2$.

However, as one introduces the interaction term $H_{\mathrm{h}}=U\sum_i n_{i,\uparrow} n_{i,\downarrow}$, the degeneracy in \eqref{degen} is lifted. This is easily seen at the thermodynamic limit when $\alpha_{p}$ are exactly localized at the edge. Then the action of $H_{\mathrm{h}}$ on the 4-dimensional edge Hilbert space is given by:
\begin{eqnarray} \label{deg_lift}
    H_{\mathrm{h}} \alpha^{\dagger}_{1,\uparrow}\alpha^{\dagger}_{1,\downarrow} |0\rangle = U \alpha^{\dagger}_{1,\uparrow}\alpha^{\dagger}_{1,\downarrow} |0\rangle, ~
     H_{\mathrm{h}} \alpha^{\dagger}_{2,\uparrow}\alpha^{\dagger}_{2,\downarrow} |0\rangle = U \alpha^{\dagger}_{2,\uparrow}\alpha^{\dagger}_{2,\downarrow} |0\rangle, ~
      H_{\mathrm{h}} \alpha^{\dagger}_{1,\uparrow}\alpha^{\dagger}_{2,\downarrow} |0\rangle = 0= H_{\mathrm{h}} \alpha^{\dagger}_{2,\uparrow}\alpha^{\dagger}_{1,\downarrow} |0\rangle
\end{eqnarray}
Hence, the 4-fold degeneracy of the edge Hilbert space is broken into two eigen-subspaces of $H_{\mathrm{h}}$, each of dimension 2. This means that the previous edge Hilbert space, which was the orbit of $SU(4)$, is now the orbit of $SU(2) \times SU(2)$. As is clear from \eqref{deg_lift}, for $U>0$, the perturbed ground state gets a contribution from the following perturbed edge state:
\begin{eqnarray}
    \tilde{|\mathrm{edge} \rangle} = \left(A_1 \alpha^{\dagger}_{1,\uparrow}\alpha^{\dagger}_{2,\downarrow} +A_2 \alpha^{\dagger}_{2,\uparrow}\alpha^{\dagger}_{1,\downarrow} \right) |0\rangle
\end{eqnarray}
This is maximally entangled when $|A_1| = |A_2| =1/\sqrt{2}$, and forms a single Bell pair state, with entanglement entropy $\ln 2$.

This analysis indicates that the drastic reduction of edge entanglement is because the inclusion of finite $U$ makes the symmetry reduction $SU(4) \rightarrow SU(2) \times SU(2)$. 
\begin{figure}[ht]
    \centering
    \includegraphics[width=0.4\textwidth]{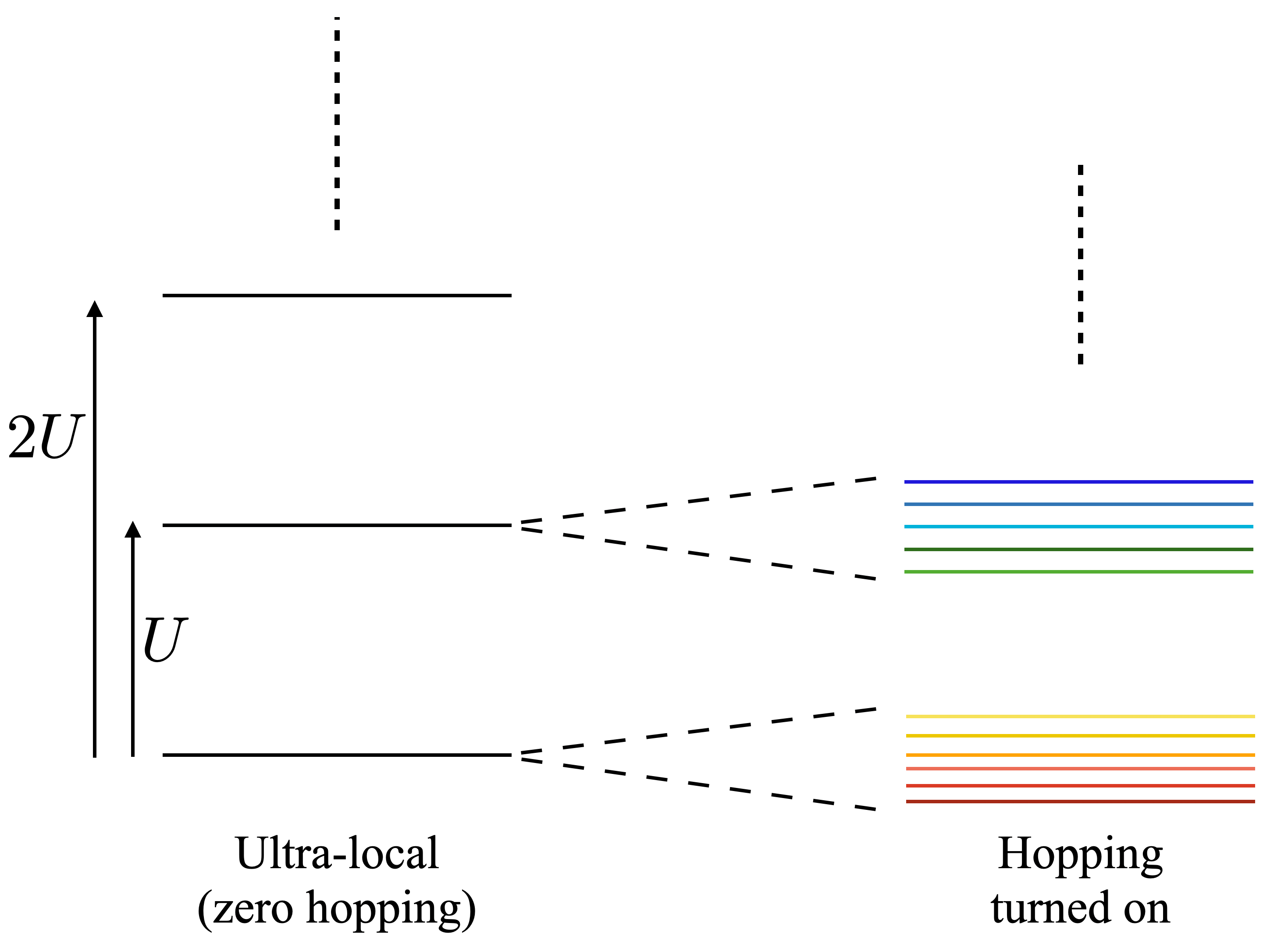}
    \caption{\textit{Left-panel:} Degenerate sub-spaces of whole state space, in the absence of hoping terms. The lowest lying line is $3^L$ times degenerate and forms the zero-energy sub-space $\mathcal{H}_\mathrm{ground}$. \textit{Right-panel:} When small hopping terms are included perturbatively, degeneracies get lifted.}
    \label{level_splitting}
    \end{figure}
\section{Entanglement signatures at extreme coupling} \label{remnant}
In this section, for convenience, we consider a single SSH %\rnote{Don't we have these results for a double chain?}
chain and study the behaviour of von Neumann entropy for open boundary conditions for very large interaction strength  ($U \gg t_2,t_1$). The following discussion, however, easily generalizes to the coupled system of chains in AB stacking as well, a point which we'll revisit while presenting numerical results.

For very strong coupling, when we can neglect the quadratic hopping terms, the Hamiltonian is just the Hubbard quartic term:
\begin{eqnarray}
    H = U \sum_i c^{\dagger}_{i,\uparrow}c_{i,\uparrow}c^{\dagger}_{i,\downarrow}c_{i,\downarrow}. \label{hub_disc}
\end{eqnarray}
This, on its own, is an ultra-local theory, described in the continuum limit by the following action:
\begin{eqnarray}
    S= \int dx dt  \left(\sum_{\sigma}i\psi^\dagger_{\sigma} \partial_t \psi_{\sigma} - U\psi^{\dagger}_{\uparrow}\psi_{\uparrow}\psi^{\dagger}_{\downarrow}\psi_{\downarrow}\right) \label{action}
\end{eqnarray}
In the continuum description, the fermionic field $\psi$ has mass dimension -1/2 and hence $U$ is dimensionless, making the theory scale-invariant. The ultra-local theory enjoys arbitrary space-dependent time-translation symmetries:
\begin{eqnarray}
    \delta_{f} \psi_{\sigma} (x,t) = f(x) \partial_{t} \psi_{\sigma},
\end{eqnarray}
for any function $f$. As a result, there are an infinite number of conserved operators, which all mutually commute:
\begin{eqnarray}
     Q_{f} = U\int dx f(x) \psi^{\dagger}_{\uparrow}\psi_{\uparrow}\psi^{\dagger}_{\downarrow}\psi_{\downarrow}, ~  [Q_{f_1},Q_{f_2}] = 0.
\end{eqnarray}
Choosing a set of functions $f$ as Dirac delta functions supported at each point on the spatial line, we see that we have a conserved operator for each degree of freedom located at each spatial point, which mutually commutes. In the discretized version, these operators are $U\psi^{\dagger}_{\uparrow}\psi_{\uparrow}\psi^{\dagger}_{\downarrow}\psi_{\downarrow}$, ie the Hamiltonian density at each point. Hence, the system is exactly solvable. In fact, in this model, the local $U(1)$ phase transformations:
$$ \psi_{\sigma} (x,t) \rightarrow e^{i \lambda (x)} \psi_{\sigma} (x,t)$$ are physical symmetries too, resulting in an infinite number of conserved quantities $P_{\lambda, \sigma} = U\int dx \lambda(x) \psi^{\dagger}_{\sigma}\psi_{\sigma}$. Just as before, one can choose $\lambda$s to be locally supported functions to construct local conserved operators. In the discretized picture, they give individual particle number operators for each site for each spin. 

As this is an exactly solvable (if not integrable) model, one can trivially construct the full spectrum of the Hamiltonian \eqref{hub_disc} or its continuum version, as this is diagonal in local Hilbert space/ computational basis. The ground state, however, is highly degenerate. This can be understood as the following. The Hamiltonian density is ultra-local and positive definite. Hence $H |\mathrm{ground} \rangle = 0$ implies 
\begin{equation}\label{constraint}
c^{\dagger}_{i,\uparrow}c_{i,\uparrow}c^{\dagger}_{i,\downarrow}c_{i,\downarrow} |\mathrm{ground} \rangle \equiv n_{i,\uparrow} n_{i,\downarrow} |\mathrm{ground} \rangle =0
\end{equation}
 for all $i$. For example, in an $L$-site system, any state with $p$ number of sites occupied by spin $\uparrow$ particles and $q \leq L-p$ other sites occupied by $\downarrow$ satisfies this condition. The dimension of this 0 energy sub-space is found to be $\sum_{p=0}^L 2^{L-p}\, {}^L C_p = 3^L$. This means that the zero-temperature thermodynamic entropy is well-defined and extensive in nature, as opposed to ultra-local bosonic models, where the canonical partition function is ill-defined \cite{de2023carroll}.  \cite{mondal2023statistical}. We name this Hilbert sub-space as $\mathcal{H}_\mathrm{ground}$. 

However, the ground state being so much degenerate, the meaning of bi-partition entanglement becomes ambiguous to a large extent. For example, the vacuum state, which also is a member of the 0 energy subspace, is entanglement-free. This is a prototypical example of an entanglement-free ground state of ultra-local conformal theories \cite{bagchi2015entanglement}, at least in cases where the ground state is unique. On the other hand, one can construct a linear combination of ${}^L C_{L/2}$ half-filled states of both spins (provided the condition \eqref{constraint} is satisfied), which is entangled, with the construction of a maximum number of Bell pairs. One of the ways to remove this ambiguity in $\mathcal{H}_\mathrm{ground}$ is by specifying all the quantum numbers of a state. Interestingly, all the quantum numbers are furnished by the local occupation numbers, i.e. $n_{i,\uparrow} n_{i,\downarrow}$, thanks to the local $U(1)$ charges. Specifying all the local site occupation numbers for each spin means fixing a particular computational basis state, which,  by definition, is entanglement-free.

Everything exactly being solvable in the ultra-local model above makes it rather uninteresting. Hence, it would be curious to ask about the fate of entanglement if one turns on at least nearest neighbour hopping terms in the Hamiltonian perturbatively. The degeneracy of the $3^L$ dimensional 0 energy sub-space now gets lifted, as presented in the cartoon Figure \ref{level_splitting}. This is because neither the local $U(1)$ symmetry nor the supertranslation charges described above are any more there in the presence of terms like $\sum_{i,j, \sigma} t_{ij} c^{\dagger}_{i,\sigma} c_{j,\sigma}$. Only the $SU(2) \times SU(2)$ global symmetry of the Hubbard model is intact. This results in the conservation of only the two quantum numbers: the total number of fermions of each spin type. Now, we are in the position of asking the question: \textit{what is the ground state of the perturbed system in the computational basis?}

\begin{figure}[ht]
    \centering
    \includegraphics[width=0.9\textwidth]{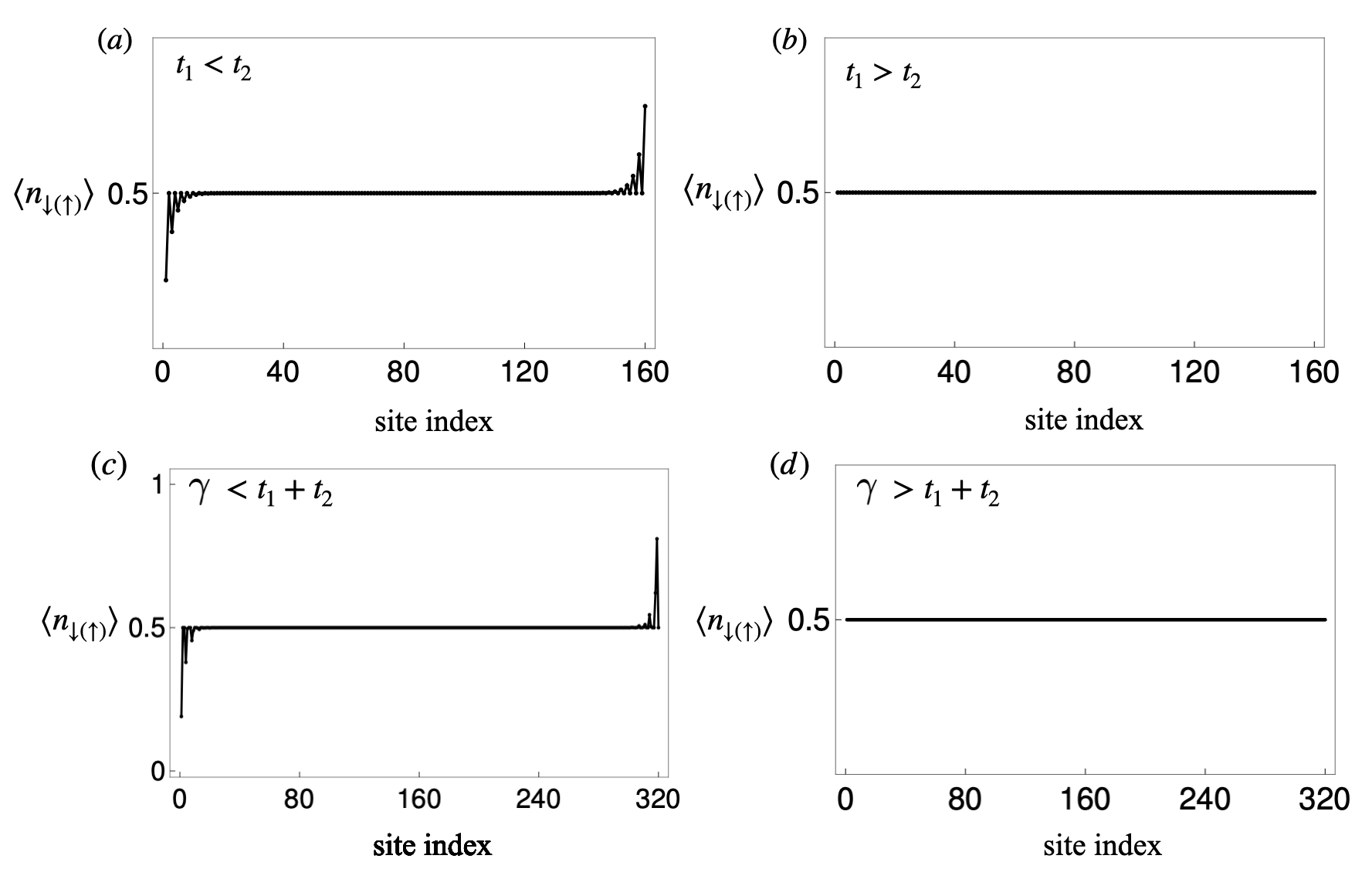}
    \caption{On-site particle densities for single SSH chain $L=160$ with $U=100$, $(a)~ t_1=0.8$ and $t_2=1.2$,  $(b)~ t_2=0.8$ and $t_1=1.2$ $(c)$;  AB stacking $L=320$ with $U=40,~t_1=0.8,~t_2=1.6$, $(c) ~\gamma=1,~(d) ~\text{and} ~~\gamma=3$. }
    
    %\rnote{@Nisa, take the Mathematica files from Indrakshi. She has fitted these data and made these graphs smooth}
    %\textcolor{blue}{(spin density calculations for AB are running.)}}
    \label{spin}
\end{figure}

To answer this question, we first notice that at the lowest order in degenerate perturbation theory, each lowest band (refer to the Figure \ref{level_splitting} right panel) energy eigenstates do belong to $\mathcal{H}_\mathrm{ground}$ and still has the constraint $n_{i,\uparrow} n_{i,\downarrow} =0, \forall i$ satisfied. We can expect that the ground state is still half-filled, as the half-filling condition doesn't violate the above constraint. To verify this, we performed DMRG calculations for the single chain and the coupled chain system in AB stacking. The parameters' details are furnished in Fig. \ref{spin}. We kept the hopping parameters well below $0.1 U$ so that the ultra-locality is only violated perturbatively. Indeed, we checked $\sum_i n^{\sigma}_{i} = L/2$ for the ground state.

\begin{figure}[h!]
    \centering
    \includegraphics[width=0.7\textwidth]{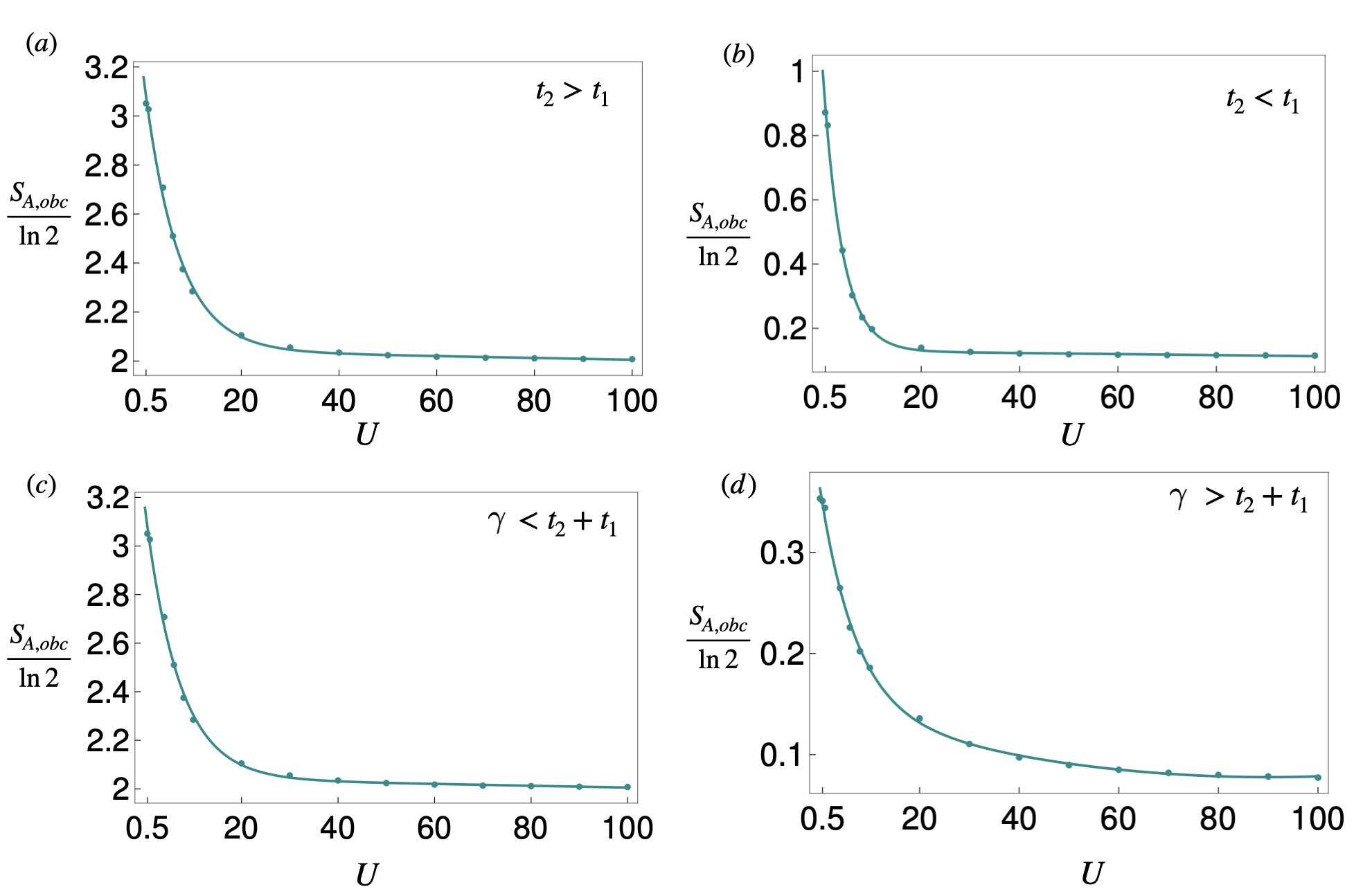}
    \caption{Remannt entropy for single chain $S_{A,obc}$ for $L=40$, $(a)~ t_1=0.8$ and $t_2=1.2$,  $(b)~ t_2=1.2$ and $t_1=0.8$; AB stacking $L=40,~t_1=0.8,~ t_2=1.6$ (a) $\gamma=1$ (b) $\gamma=3$
    }
    %\rnote{@Nisa, take the Mathematica files from Indrakshi. She has fitted these data and made these graphs smooth}}
    \label{rs}
\end{figure}

The next level of the query is definitely the bi-partite entanglement of the ground state. However, it is subject to the constraint that the zero-hopping ground state space had: $n_{i,\uparrow} n_{i,\downarrow} = 0$. This means once we choose any of the ${}^{L}C_{L/2}$ half-filled states for up-spins, the down-spins do have only one way to fill half of the lattice sites. In other words, the ground state now is a linear combination of ${}^{L}C_{L/2}$ computational basis states \footnote{This argument is valid only in the lowest order in perturbation theory of hopping parameters. However, this being strong coupling perturbation theory, the perturbative correction to the states in $\mathcal{H}_{\mathrm{ground}}$ received from excited states are doubly suppressed by a factor of $\sim \delta t/U$. }. This contrasts what happens when we have a free theory, ie. $U=0$, and the ground state is a linear combination of  $({}^{L}C_{L/2})^2$ computational basis states. Hence, a drastic drop (almost to zero) in entanglement for a large $U$ regime is expected. DMRG computation validates this argument, too, as seen in panels (b) and (d) of figure \ref{rs}, respectively, for a single SSH and the coupled system in AB stacking. For the DMRG computations, we present here only the OBC case. For large $U \gg t_1, t_2$, all the arguments mentioned above stand both for OBC and PBC alike. However, the vanishing of entanglement holds only if edge effects are not considered, which are substantial in the OBC. As shown in the plots \ref{spin}, even in this very strong coupling regime small values of the hopping parameters still result in non-trivial edge localization in the ground state for the topologically non-trivial phase. There is a clear edge occupancy imbalance in these phases, whereas the ground state has exact half-filling in each site for the trivial phase. This non-trivial edge effect is manifest in the edge entanglement measuring $2 \ln 2$ in both the single SSH and the coupled chain system, as evident in panels (a) and (c) of Fig. \ref{rs}. This is what we name the remnant entropy, i.e. \textit{the one remnant even in the strongly coupled ultra-local theory, perturbed by a small but topologically non-trivial hopping term}.
%For the non-trivial phase with zero energy edge modes, the $S_{A,obc}$ is \textit{remnant entropy} as $2\ln{2}$ as shown in Fig.~\ref{rs}. This could be explained as the following: consider the system with onsite Hubbard interactions only. In that case, with all the quartic terms, the Hamiltonian is diagonal. The ground state is the vacuum state $\ket{\text{ground}}=\ket{0}\otimes~ ...~ \otimes \ket{0},\ $ which is a tensor product state, hence, $S_{A,obc}=0$. With large $U$, the nearest hopping parameter could be treated as a perturbation. For $t_1<t_2$, edge entanglement is restored and is 0 otherwise. 
%\textcolor{blue}{AB calculations are not complete yet}

%\subsection{Central Charge}
%{\textbf{Better result}}

%\rnote{Put the following in Appendix}
\section{Conclusion and Outlook}\label{5}

In this work, a bi-layer SSH model with and without on-site Hubbard interaction is explored in light of the entanglement properties of the system. As mentioned earlier, for free theory, i.e. when the Hubbard interaction is absent, there is good analytic control over the model, and one can calculate entanglement entropies from the correlation matrix of the theory semi-analytically. With a proper definition of edge entanglement, one can find a clear signature of the presence of topological order in the system, establishing the edge entanglement as an order parameter for the quantum phase transition. Within this study, we have even found that edge contribution calculated from the symmetry-resolved entanglement for the system does qualify as an order parameter. The concept of equipartition of entanglement and the study of symmetry-resolved entanglement are relatively recent concepts, and the power of these in systems with global symmetries is yet to be fully explored. Our analysis demonstrates a clear application for studying symmetry-resolved entanglement. Instead of studying the entire system, focusing only on a particular global symmetry sector is computationally cheap and would allow one to invade a class of theories which were computationally intractable otherwise. The impact of symmetry-resolved entanglement for an interacting theory and a theory containing non-Abelian anyons would be of extreme interest to the community.

Moving beyond the free theory, analytic understanding of the system is limited. Studying the spectrum and its entanglement structure for such a system is extremely expensive in terms of classical computational resources. In this work, we utilize state-of-the-art Hamiltonian simulation algorithms via the tensor network ansatz for the states and operators of the system. Note that tensor network algorithms may not be the most efficient ones. Still, they are useful tools for studying the low energy spectrum for reasonably large systems on a classical computer, which is also essential for benchmarking quantum algorithms. Using tensor network algorithms for a fermionic system and for a system with periodic boundary conditions are particularly nontrivial, albeit essential for this project. Benchmarking the tensor network codes by reproducing the exact analytic results for a free theory is crucial and has been implemented in this work. This study establishes the power of DMRG calculation to be used in the future for more complicated systems with a rich symmetry structure. Note that the DMRG algorithms, being governed by the study of symmetry-resolved entanglement, can access much larger system size and do provide signatures of different topological phases, and the result is reliable as established in section \ref{2}. Using tensor network algorithms, this work establishes the edge entanglement to be an order parameter for the topological phase transition in the case of an interacting theory.

Studying an interacting theory is of utmost interest as it mimics real physical systems more reliably. One important finding of this work is the notion of a remnant entanglement in the topological nontrivial phase, which remains protected for arbitrarily large value of the interaction strength. This denotes the edge entanglement to remain protected under strong correlation. This fact might impact engineering stable correlated qubits using real physical systems.

In this work, the study of the remnant entropy for an interacting system led to a fascinating understanding of the basic structure of the low-energy Hilbert spaces. It is well understood that for arbitrary large strength of Hubbard interaction, the energy eigenstates of the ultralocal Hamiltonian are not entangled. However, even if there is a hopping present in the perturbative regime, this picture changes and the ground state acquires a quantized amount of entanglement entropy, which we refer to as the remnant entanglement and is robust under any change of the parameter values, as long as the system remains in the non-trivial topological phase. The study of the ground state configuration for interacting theories does help in revealing the mystery, as discussed in detail in the last part of section \ref{5}. Future research would definitely explore this in detail for more complicated models, if possible in higher dimensions. We expect the upcoming era of quantum computation will be able to shed some light on understanding the structure of the occupation number basis in real space and its sub-structures. Studying the fragmentation of the Hilbert space is important for choosing a suitable sector to work on with limited computational resources as well as for understanding underlying global symmetries for each sector.

\section*{Acknowledgements}
Research of RB is supported by the following grants from the SERB, India: CRG/2020/002035, SRG/2020/001037, MTR/2022/000795 and OPERA grant from BITS Pilani. Research of IR is supported by the  OPERA award (FR/SCM/11-Dec-2020/PHY) from BITS-Pilani, the Start-up Research Grant (SRG/2022/000972) and Cor-Research Grant (CRG/2022/007312) from SERB, India and the cross-discipline research fund (C1/23/185) from BITS Pilani. We thank Aritra Banerjee, Arjun Bagchi, Arpan Bhattacharyya, Arti Garg, and Krishanu Roychowdhury for insightful discussions. NA acknowledges the hospitality of ICTS-TIFR and NISER during the final stages of writing
the draft.  

\appendix
\section*{Appendix}
\subsection{System size effects}\label{a1}
\begin{figure}[ht]
    \centering
    \includegraphics[width=0.7\textwidth]{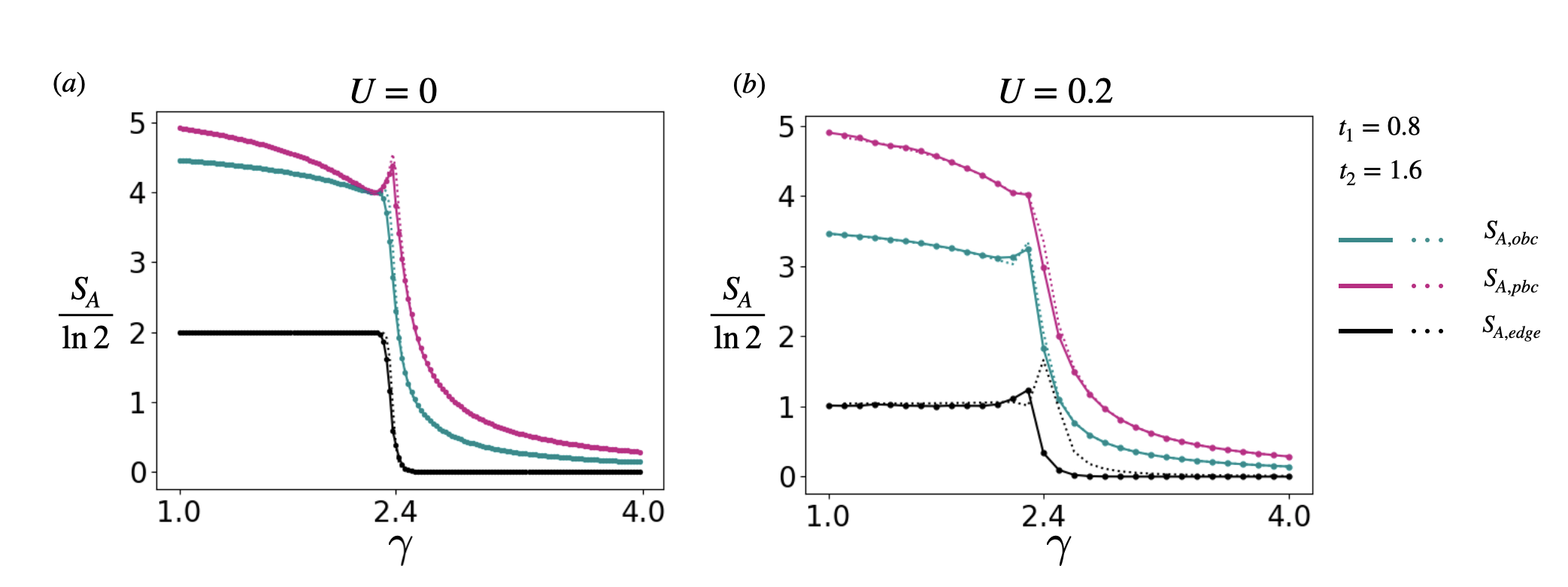}
    \caption{Solid lines show subsystem entanglement for $L=40$ and dotted lines near $\gamma=2.4$ are for $L=56$. $S_{edgde}$ remains a topological invariant as $(a)$  $2\ln{2}$ for free full theory and $(b)$ $\ln{2}$ for interacting theory independent of system size.}
    \label{size}
\end{figure}

In order to see how $S_{edge}$ changes with system size, we calculate subsystem entanglement for $L=56$ and compare it with that for $L=40$. From Fig.~\ref{size}, what could be inferred is that $S_A$ does not change with the size of the system ($L$) away from the critical point for open and periodic boundary conditions.

\label{DMRGdetails}
\bibliography{ref}
\end{document}